\documentclass[11pt,twocolumn]{IEEEtran}
\usepackage{lipsum}

\usepackage{cite}
\ifCLASSINFOpdf
   \usepackage[pdftex]{graphicx}
\else
\fi
\usepackage[cmex10]{amsmath}
\usepackage{amsfonts}
\usepackage{algorithmic}
\usepackage{array}
  \usepackage[caption=false,font=normalsize,labelfont=sf,textfont=sf]{subfig}
\usepackage{fixltx2e}
 \usepackage{dblfloatfix}
\ifCLASSOPTIONcaptionsoff
  \usepackage[nomarkers]{endfloat}
 \let\MYoriglatexcaption\caption
 \renewcommand{\caption}[2][\relax]{\MYoriglatexcaption[#2]{#2}}
\fi
 \let\MYorigsubfloat\subfloat
 \renewcommand{\subfloat}[2][\relax]{\MYorigsubfloat[]{#2}}
\usepackage{url}
\begin{document}
\title{A Convex Optimization Approach to pMRI Reconstruction}

\author{Cishen~Zhang and
        Ifat~Al~Baqee% <-this % stops a space
\thanks{Faculty of Engineering and Industrial Sciences,
Swinburne University of Technology,
Hawthorn, VIC 3122, Australia, e-mail: (cishenzhang@swin.edu.au and ibaqee@swin.edu.au).}% <-this % stops a space
%\thanks{Manuscript received April 19, 2005; revised December 27, 2012.}
}

\maketitle

\begin{abstract}
In parallel magnetic resonance imaging (pMRI) reconstruction without using pre-estimation of coil sensitivity functions, one group of algorithms reconstructs sensitivity encoded images of the coils first followed by the magnitude image reconstruction, e.g. GRAPPA. Another group of algorithms jointly computes the image and sensitivity functions by regularized optimization which is a non-convex problem with local only solution. For the magnitude image reconstruction, this paper derives a reconstruction formulation, which is linear in the magnitude image, and an associated convex hull in the solution space of the formulated equation containing the magnitude image. As a result, the magnitude image reconstruction for pMRI is formulated into a two-step convex optimization problem, which produces a globally optimal solution. An algorithm based on split-bregman and nuclear norm regularized optimizations is proposed to implement the two-step convex optimization and its applications to phantom and in-vivo MRI data sets result in superior reconstruction performance compared with existing algorithms.
\end{abstract}


% \fi
\IEEEpeerreviewmaketitle

\section{Introduction}
Magnetic resonance imaging (MRI) is an advanced modality for noninvasive medical diagnosis with continuously growing clinical applications. In comparison with other medical imaging modalities, such as x-ray computed tomography (CT) and ultrasound, MRI is beneficiary in a way that it provides very safe scanning, high spatial resolution and flexible contrast for displaying body tissues. It is, however, also known that the data acquisition of MRI is a relatively long process, which is governed by the time required for physical excitation and relaxation of the radio frequency (RF) magnetic field. The relatively low speed of MRI scan can be an uncomfortable experience for patients and can result in low patient throughput of MRI scan operation. It can also cause motion artifacts in images, typically for respiratory or cardiac organs which have motions during the examination.

To accelerate the scan speed without compromising the image quality has been an important and challenging problem in the MRI research. For this purpose, modern MRI scanners implement multiple receiver coils in phased array mode for parallel acquisition of MRI data in the $k$-space. This technology is known as parallel MRI (pMRI). In pMRI, distinct spatial sensitivities of the receiver coils can enable simultaneous acquisition of $k$-space data containing complementary information.
As a result, the pMRI can accelerate MRI scans to considerably reduce the overall scan time and a combined set of undersampled partial $k$-space data acquired from different receiver coils can provide sufficient information for image reconstruction.

The MRI produces gray value images displaying the spatial magnetic spin density function of the imaged object. The magnetized spin density function is complex valued with magnitude and phase and is determined by the proton density of the imaged object, the external magnetic field and RF excitation pulses. The coil sensitivity functions are also complex valued and bounded due to their finite inductance values \cite{roemer_1990}. To reconstruct the MR image from the $k$-space data of a single receiver coil scanner is an inverse Fourier transform process under the uniform sensitivity assumption. But the pMRI reconstruction using undersampled $k$-space data is not a straightforward task. It requires knowledge of spatial sensitivity functions of the multiple receiver coils, which are in general not only determined by the coil instrumentation but also dependent upon the imaged object. There have been numerous algorithms developed in past years for pMRI reconstruction. Depending on how the information of sensitivity functions is incorporated into the image reconstruction and how the image information is reconstructed, the existing reconstruction algorithms can be classified into three groups.

The first group of algorithms pre-estimates the complex valued sensitivity functions and use the estimated results to reconstruct the magnitude and phase functions of the complex valued image. The performance of the algorithms depends on the accuracy of the pre-estimated sensitivity functions. Typical algorithms of this group are SMASH\cite{sodik_1997}, SENSE \cite{pruessmann_1999}, and their extensions such as \cite{kyria_2000, liu_2007, Madore_2004}. Also included in this group are some algorithms based on estimation of sensitivity functions and regularized optimization e.g. \cite{majumdar_2012, chen_2012, liu_2008, liang_2009, huang_2010}. The second group of algorithms estimates the sensitivity encoded images of each receiver coil first followed by a image reconstruction operation to obtain the image. These algorithms do not require knowledge of the sensitivity functions but the reconstructed image function is magnitude only without containing the phase information. Typical algorithms of this group is GRAPPA \cite{griswold_2002}, IIR GRAPPA \cite{zhaolin_2010} and their extensions using the sum-of-squares (SOS) operation \cite{roemer_1990}. There are also recent algorithms which reconstruct the sensitivity encoded images by regularized optimization, e.g. \cite{lustig_2010, murphy_2012, weller_2013, park_2012}. The third group of algorithms formulates the pMRI reconstruction into a regularized optimization problem without requiring estimation of sensitivity functions \cite{leslie_2007,uecker_2008,huajun_2010,derya_2011,knoll_2012}. The algorithms jointly compute the complex valued image and sensitivity functions by minimizing a performance index function which incorporates the reconstruction error and regularization terms. Because of the inherent cross product terms of the image and sensitivity functions, the formulated optimization problem is bilinear in the optimization variables and hence nonconvex. It therefore can only result in a local solution depending on the selection of initial condition and has difficulties in finding the global optimal solution as well as computational complexity.

This paper considers the pMRI reconstruction problem by regularized optimization without using knowledge of the sensitivity functions and tackles the difficulties of nonconvex optimization and local solution of the third group of algorithms. It is shown that, if only the magnitudes of the image and sensitivity functions are reconstructed, the pMRI reconstruction can be formulated into a linear and convex optimization problem which has a global optimal solution. The linear and convex formulation of the problem can lead to efficient computing of the solution and the global optimal solution can outperform other pMRI reconstruction algorithms. Without loss of popularity and as the second group of algorithms including GRAPPA and its extensions have done, the magnitude only image reconstruction can meet the needs of most clinical applications. It is also noted that there are some application cases, such as phase contrast imaging for detection of the velocity of flow \cite{pelc_1991}, where the phase infromation of the image is required and the magnitude only image reconstruction is not sufficient.

Like the two-step procedure of GRAPPA and its extensions, which first estimate the sensitivity encoded image functions of each coil followed by an SOS operation to construct the magnitude image, the proposed pMRI reconstruction in this paper is formulated into a two-step convex optimization problem, with the first step optimization solving the sensitivity encoded image functions of each coil and the second step optimization solving the magnitude image function. The two optimization steps operate sequentially in a way that the second step optimization is carried out after completion of the first step optimization, which is different from the iterative alternating optimization for solving the nonlinear optimization problems. The two-step convex optimization is implemented with an algorithm based on Split-bregman method\cite{goldstein_2009} and nuclear norm regularization\cite{majumdar_2011} and applied to in-vivo $k$-space data for pMRI reconstruction. Its performance in reconstruction accuracy and efficiency in comparison with other methods is demonstrated.

In this paper, $\mathbb{R}$, $\mathbb{R}_+$ and $\mathbb{C}$ denote the sets of real, nonnegative real and complex numbers, respectively. The lower bold case letter denotes vectors and the capital bold case letter denotes matrices. $\preceq$ and $\succeq$ denote the elementwise operations of $\leq$ and $\geq$ on vectors, respectively. $\odot$ denotes the Hadamard or elementwise product of vectors. $| \cdot |$ takes
elementwise magnitude of vectors and $\angle$ unitizes elements of vectors, such that $\mathbf{v} = |\mathbf{v}|\odot \angle \mathbf{v}$ for a complex valued vector $\mathbf{v}$.
$\langle \cdot,\cdot \rangle$ denotes the inner product of vectors. $\mathbf r = (x,y)$ and $\mathbf k=(k_x,k_y)$ denote the 2D coordinate systems of the spatial image and $k$-space domains, respectively.
\section{Formulation of the $k$-space data for convex optimization}
\subsection{The undersampled $k$ space data}
Consider a pMRI scanner implemented with $L$ receiver coils. Let $h(\mathbf{r})\in \mathbb{C}$ be the 2D spatial MR image function and $s_i(\mathbf{r}) \in \mathbb{C}$, $i=1,\cdots, L$, be the 2D spatial sensitivity functions of the coils. The sensitivity encoded image functions of the coils are $z_i(\mathbf{r})=h(\mathbf{r})s_i(\mathbf{r}) \in \mathbb{C}$, $i=1,\cdots, L$, which are products of $h(\mathbf{r})$ and $s_i(\mathbf{r})$. The MRI scan creates the following $k$-space functions of the $L$ receiver coils
\begin{equation}\label{model1}
  g_i(\mathbf{k})=  \int \int z_i(\mathbf{r})e^{-j2\pi\langle\mathbf{k},\mathbf{r}\rangle}d\mathbf{r},\quad i=1,\cdots,L,
\end{equation}
which are the Fourier transforms of $z_i(\mathbf{r})=h(\mathbf{r})s_i(\mathbf{r})$.

The discrete version of the Fourier transform equations in
(\ref{model1}) can be written in the vector forms as
\begin{equation}\label{dft}
\mathbf g_i =\mathbf{F} \mathbf{z}_i = \mathbf{F} (\mathbf{s}_i\odot\mathbf{h}), \ \ i=1,2,\cdots,L,
\end{equation}
where $\mathbf{h}, \ \mathbf{s_i}, \ \mathbf{z_i}, \ \mathbf{g_i} \in \mathbb{C}^{N^2}$, $i=1,\cdots, L$, are the discretized vectors of $h(\mathbf{r})$, $s_i(\mathbf{r})$, $z_i(\mathbf{r})$ and $g_i(\mathbf{k})$,
respectively, and the matrix $\mathbf{F} \in \mathbb{C}^{N^2\times N^2}$ operates the 2D discrete Fourier transform (DFT) on the vectorized form of 2D matrices. The undersampled vectors of the $k$-space data $\mathbf g_i$, denoted by $\tilde {\mathbf g}_i \in \mathbb{C}^M $, with $M < N^2$, can be represented as
\begin{equation}\label{undersampled}
\tilde {\mathbf g}_i = \tilde {\mathbf F} \mathbf{z }_i = \tilde {\mathbf F} (\mathbf{s}_i\odot\mathbf{h}), \ \ i=1,2,\cdots,L,
\end{equation}
where $\tilde {\mathbf F} \in \mathbb{C}^{M\times N^2}$ is the corresponding undersampled
2D DFT matrix operating on vectors. 
\subsection{The convex solution space}
Given the undersampled $k$-space data vectors $\tilde {\mathbf g}_i$ in the form (\ref{undersampled}), to find a joint solution for the image ${\mathbf h}$ and sensitivity functions ${\mathbf s}_i$ is in general a nonlinear and nonconvex problem. If only the magnitude of the image function is considered, it is possible to construct a convex solution space for the magnitude image and the sensitivity encoded functions ${\mathbf z}_i$, which can further lead to a convex optimization formulation of the image reconstruction. An intuitive observation of the convex solution space is provided below.

Let ${\mathbf h}_m = \in \mathbb{R}_+^{N^2}$ be the magnitude of the image vector ${\mathbf h}$. Since the magnitudes of ${\mathbf s}_i$ are bounded due to bounded inductances of the coils, there exist constant vectors ${\mathbf b}_i \in \mathbb{R}_+^{N^2}$ such that $|{\mathbf s}_i| \preceq {\mathbf b}_i$, $i=1, \cdots, L$. It follows that
\begin{equation}\label{bounds}
|{\mathbf z}_i| \preceq \mathbf{b}_i \odot \mathbf{h}_m, \ \ i=1,\cdots,L.
\end{equation}

\begin{figure}[!t]
\centerline
{\includegraphics[width=3in]{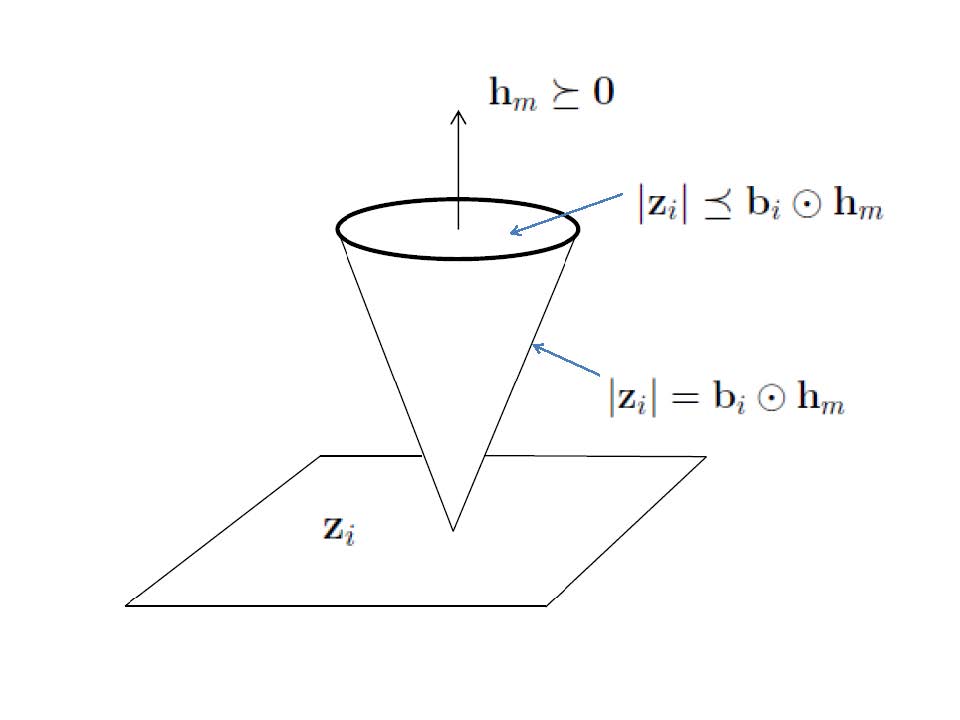}}
\caption{Convex solution space for ${\mathbf h}_m$ and ${\mathbf z}_i$. }
\label{cone}
\end{figure}

In each bilinear equation ${\mathbf z}_i=\mathbf{s}_i \odot \mathbf{h}_m$ of the sensitivity encoded image functions, there are two independent variable vectors which, if known, can determine the third vector variable. If ${\mathbf h}_m$ and ${\mathbf z}_i$ are considered as the solution variables, the inequalities (\ref{bounds}) form a cone shaped convex hull containing the solutions of ${\mathbf h}_m$ and ${\mathbf z}_i$, with properly chosen constant bound vectors ${\mathbf b}_i$. Such a convex solution space is
displayed in Fig. \ref{cone}, on top of the complex plane of ${\mathbf z}_i$, for the scalar case of ${\mathbf h}_m \in \mathbb{R}_+$ and ${\mathbf z}_i\in \mathbb{C}$. This solution space provides a basis for the convex optimization of the pMRI reconstruction problem and its extension to the high dimensional convex solution space is straightforward. It is, however, noted that
the convex solution space only exists for the magnitude image ${\mathbf h}_m$ but not for any other real or complex valued image vectors.
\subsection{Linear formulation of the pMRI reconstruction}
The solution space for ${\mathbf h}_m$ and ${\mathbf z}_i$, as shown in Fig.\ref{cone}, displays the 
convex nature of the problem. But, as seen from (\ref{undersampled}), the magnitude image ${\mathbf h}_m$ is a bilinear variable, coupled with ${\mathbf s}_i$, of the composite vectors ${\mathbf z}_i$, $i=1,\cdots,L$, so is not a linear variable of the problem equation.
To facilitate formulation of a linear model for the convex optimization, introduce the magnitude and phase vectors of $\mathbf{z}_i$ denoted by $\mathbf{m}_i=|\mathbf{s}_i\odot\mathbf{h}|$ and $\mathbf{p}_i=\angle (\mathbf{s}_i \odot \mathbf{h})$, respectively, such that $\mathbf{z}_i=\mathbf{m}_i \odot \mathbf{p}_i$, $i=1,\cdots,L$. Further introduce decoupling parameter vectors ${\mathbf d}_i \in \mathbb{R}_+^{N^2}$, $i=1,\cdots, L$, and denote their corresponding diagonal matrices as ${\mathbf D}_i$. Using ${\mathbf d}_i$, the magnitude vectors ${\mathbf m}_i$ can be written as
\begin{equation}\label{mi}
{\mathbf m}_i = -{\mathbf d}_i \odot {\mathbf h}_m  + (|{\mathbf s}_i|+{\mathbf d}_i)\odot {\mathbf h}_m
 = [-{\mathbf D}_i \ \ {\mathbf I}]
\left[\begin{array}{c} {\mathbf h}_m \\
{\mathbf {\bar m}}_i \end{array}\right], 
\end{equation}
$i=1,\cdots, L$, where ${\mathbf {\bar m}}_i= (|{\mathbf s}_i|+{\mathbf d}_i)\odot {\mathbf h}_m$. It follows from $|{\mathbf s}_i| \preceq {\mathbf b}_i$ that ${\mathbf {\bar m}}_i$ are linearly and elementwisely bounded by ${\mathbf h}_m$, i.e.
\begin{equation}\label{bmi}
{\mathbf {\bar m}}_i \preceq {\mathbf c}_i \odot {\mathbf h}_m, \ \
i=1,\cdots, L
\end{equation}
where ${\mathbf c}_i={\mathbf b}_i+{\mathbf d}_i$.

Let $\tilde {\mathbf g}\in \mathbb{C}^{LM}$, ${\mathbf z}, {\mathbf p} \in \mathbb{C}^{LN^2}$ and ${\mathbf m},  \bar {\mathbf m} \in \mathbb{R}_+^{LN^2}$ be
the stacked vectors of ${\mathbf g}_i, \ {\mathbf z}_i, \ {\mathbf p}_i, \ {\mathbf m}_i$ and $\bar {\mathbf m_i}$, $i=1,\cdots, L$, respectively.
The undersampled $k$-space vectors $\tilde {\mathbf g}_i$ in (\ref{undersampled}) can be rewritten as
\begin{equation}\label{tildeg}
\tilde {\mathbf g} =\mathbf{\bar F} {\mathbf z} =\mathbf{\bar F} ({\mathbf m} \odot {\mathbf p}),
\end{equation}
where $\mathbf{\bar F} \in \mathbb{C}^{LM \times LN^2}$ is the blocked diagonal matrix of $\tilde {\mathbf F}$. The stacked vector form of ${\mathbf m}_i$ in (\ref{mi}) is
\begin{equation}\label{m}
{\mathbf m} = \bar {\mathbf D}
\left[\begin{array}{c} {\mathbf h}_m \\
{\mathbf {\bar m}} \end{array}\right],
\end{equation}
where
$$\bar {\mathbf D} = \left[\begin{array}{cccc}
-{\mathbf D}_1 & & & \\
\vdots & &{\mathbf I} &  \\
-{\mathbf D}_L & & &
\end{array}\right]\in \mathbb{R}^{LM\times(L+1)N^2}.
$$
The vector equation (\ref{m}) shows that the magnitude ${\mathbf h}_m$ of the image function is linearly decoupled from the magnitude vector ${\mathbf m}$ of the sensitivity encoded image functions. This technical result is instrumental for the proposed convex optimization for pMRI reconstruction.

\section{Two-step convex optimization for pMRI reconstruction}
\subsection{General formulation}
It is observed from (\ref{tildeg}) that the global solution for the sensitivity encoded image vector ${\mathbf z}$ and hence its magnitude ${\mathbf m}$ and phase ${\mathbf p}$ can be obtained by solving the linear equation ${\mathbf {\tilde g}}={\mathbf {\bar F}} {\mathbf z}$. Using the solution
for ${\mathbf m}$, the linear equation (\ref{m}) can be further solved to obtain a solution for the magnitude ${\mathbf h}_m$ of the image. Once the magnitude ${\mathbf h}_m$ is obtained. The magnitudes $|{\mathbf s}_i|$, $i=1,\cdots, L$, of the sensitivity functions can be further determined using (\ref{mi}).
Based on this observation, the general formulation of the proposed pMRI reconstruction consists of two sequential convex optimization steps ${\cal P}_1$ and ${\cal P}_2$ as follows.

Based on [\ref{tildeg}], the first step solves the complex valued sensitivity encoded image vector ${\mathbf z}$ and hence its magnitude ${\mathbf m}$ and phase ${\mathbf p}$
by the following regularized convex optimization
\begin{equation}\label{dconvex1}
\begin{array}{l}
{\cal P}_1:
\begin{array}{l}
\min_{\mathbf{z}} {1 \over 2}\| \tilde {\mathbf{g}} - \bar {\mathbf{F}} {\mathbf{z}}\|_2^2 + R_1({\mathbf{z}}),
%\textrm{subject to: \ [\ref{tildeg}] \ and} \ \mathbf f_1(\mathbf{z})\preceq {\mathbf 0}.
\end{array}
\end{array}
\end{equation}
where $R_1({\mathbf{z}})$ is a convex regularization function to be further specified according to application conditions.

Suppose that $\mathbf{z}^o=\mathbf{m}^o \odot \mathbf{p}^o$ is the optimal solution of the optimization problem ${\cal P}_1$ with $\mathbf{m}^o = |\mathbf{z}^o| \in \mathbb{R}_+^{LN^2}$ and
$\mathbf{p}^o = \angle \mathbf{z}^o \in \mathbb{C}^{LN^2}$ being the corresponding optimal solutions of $\mathbf{m}$
and $\mathbf{p}$, respectively. Substituting $\mathbf{m}^o=\mathbf{m}$ into (\ref{m}) yields
\begin{equation}\label{mo}
{\mathbf m}^o = \bar {\mathbf D}
\left[\begin{array}{c} {\mathbf h} \\
{\mathbf {\bar m}} \end{array}\right].
\end{equation}
Given ${\mathbf m}^o$, the second step of the proposed convex optimization is
\begin{equation}\label{dconvex2}
\begin{array}{l}
{\cal P}_2:
\begin{array}{l}
\min_{\mathbf{h}_m,\mathbf{\bar m}} {1 \over 2} \left\|
\bar {\mathbf m}^o - \bar {\mathbf D} \left[\begin{array}{c} {\mathbf h} \\
{\mathbf {\bar m}} \end{array}\right] \right\|_2^2 + R_2(\mathbf{h}_m,\mathbf{\bar m}),\\
\textrm{subject to:} \ \ \mathbf{h}_m \succeq {\mathbf 0}, \
{\mathbf 0} \preceq \mathbf{\bar m} \preceq {\mathbf c} \odot {\mathbf {\bar h_m}},
\end{array}
\end{array}
\end{equation}
where $R_2(\mathbf{h}_m,\mathbf{\bar m})$ is a convex regularization function to be further specified according to application conditions, ${\mathbf {\bar{h}_m}}, {\mathbf c} \in \mathbb{R}_+^{LN^2}$ are the $L$-fold stacked vector of ${\mathbf h_m}$ and the stacked vector of ${\mathbf c}_i$, $i=1, \cdots, L$, respectively. Since the linear equation (\ref{mo}) is underdetermined and has infinite number of solutions, the inequality ${\mathbf 0} \preceq \mathbf{\bar m} \preceq {\mathbf {\bar h}_m}$ based on (\ref{bmi}) forms a convex hull to constrain ${\mathbf h}_m$ and $\bar {\mathbf m}$ in the solution space. The solution for ${\cal P}_2$ provides the optimal magnitude image $\mathbf{h}_m^o$ as well as the optimal $\bar {\mathbf{m}}^o$. The corresponding optimal solutions $|\mathbf{s}_i|^o$, $i=1,\cdots, L$, for the magnitude of sensitivity functions can be further determined by
${\mathbf{\bar m}}_i^o=
(|{\mathbf{s}}_{i}|+{\mathbf{d}}_i) \odot {\mathbf{h}}_m^o$.
At some points where the image function has zero values, feasible solution values for sensitivity functions are not available. Proper interpolation may be introduced at these points for reconstruction of the magnitude sensitivity functions based on their smooth property.

In the two optimization steps ${\cal P}_1$ and ${\cal P}_2$, ${\cal P}_1$ is to optimize the solution for (\ref{tildeg}) and is originally a linear problem. The convexity of the optimization problem ${\cal P}_2$ is built upon the decoupled linear equation (\ref{mo}) and convex solution space specified by (\ref{bmi}). This is possible only if the solution variable is the magnitude only image vector. For complex valued ${\mathbf{z}}^o$ and ${\mathbf{h}}$, although a decoupling linear equation in the same form of (\ref{mo}) can be formulated, a convex set in the solution space does not exist for the decoupled variables, so the problem remains nonlinear and nonconvex.

\subsection{Split-bregman and nuclear norm regularized optimizations}
The above formulated convex optimization steps ${\cal P}_1$ and ${\cal P}_2$ are in general forms and can be implemented with different regularity functions and variable constraints. Taking into account properties of MR images and sensitivity functions, this subsection presents an implementation of
${\cal P}_1$ with split-bregman method and the nuclear norm regularized optimization for implementation and computation of ${\cal P}_2$.

The sensitivity encoded function $\mathbf{z}$ to be optimized in problem ${\cal P}_1$ is a product of the image and sensitivity functions and typically can have a piecewise smooth characteristic. The implementation of problem ${\cal P}_1$ takes this characteristic into account and incorporates it into the regularization function $R_1(\mathbf{z})$.
It is known that Bregman iteration \cite{bregman_1967} can solve a broad class of regularized optimization problems. It can result in superior image reconstruction performance when a hybrid of Bounded-Variation (BV) and Besov regularization is used \cite{goldstein_2009} and has been applied to MR image reconstruction, e.g. \cite{liu_2009}. Applying the split-bregman method to ${\cal P}_1$,
the regularization function $R_1({\mathbf{z}})$ can be formulated as
\begin{equation}\label{j11}
 R_1({\mathbf{z}}) = {\lambda_1}\|{\mathbf{z}}\|_{BV} + {\gamma_1}\|{\mathbf{W}}{\mathbf{z}}\|_1,
\end{equation}
where $\|\cdot\|_{BV}$ denotes the bounded variation norm defined as $\|{\mathbf{z}}\|_{BV} = \sum \sqrt {|\nabla_x z|^2+|\nabla_y z|^2}$ with
$\nabla_x$ and $\nabla_y$ being the difference operators in the $x$ and $y$ directions, respectively,
${\mathbf{W}}$ is a wavelet transform matrix and $\lambda_1$ and $\gamma_1$ are regularization parameters. The regularization terms $\|{\mathbf{z}}\|_{BV}$ and $\|{\mathbf{W}}{\mathbf{z}}\|_1$ in (\ref{j11}) are to promote, respectively, the piecewise smoothness of and energy compactness of ${\mathbf{z}}$.
The two regularization terms, together with (\ref{dconvex1}), yield the following split-bregman regularized optimization for solving ${\cal P}_1$
\begin{equation}\label{opt11}
\min_{\mathbf{z}}
{1\over 2}\|\bar {\mathbf{F}}_c{\mathbf{z}}-\tilde {\mathbf{g}}\|_2^2 + {\lambda_1} \|{\mathbf{z}}\|_{BV} + {\gamma_1} \|{\mathbf{W}}{\mathbf{z}}\|_1.
\end{equation}

In the optimization of problem ${\cal P}_2$, a reduction of the magnitude ${\mathbf h}_m$ in the solution of the underdetermined linear equation (\ref{mo}) can result in the value of $\mathbf{\bar m}$ and hence the values of $|\mathbf{s}_i|$ to grow. Thus it requires a proper scale of the solutions for
${\mathbf h}_m$ and $\bar {\mathbf m}$ by the regularization function $R_2({\mathbf h}_m,\bar {\mathbf m})$ and appropriate constraints on the solutions. The implementation of ${\cal P}_2$ considers the nuclear norm regularized optimization \cite{recht_2010} which has shown promising results in computational efficiency and accuracy in the application to MR image reconstruction \cite{majumdar_2011}, \cite{majumdar_2012}. The nuclear norm of the magnitude image vector, denoted by $\|{\mathbf h}_m\|_\ast$, is defined as the sum of singular values of
${\mathbf h}_m$. To use $\|{\mathbf h}_m\|_\ast$ as a regularization term together with the inequalities [\ref{bmi}], which specify that the solution for $\mathbf{\bar m}$ is linearly bounded by ${\mathbf h}_m$ in a convex hull, can provide proper scaling and effective constraints on ${\mathbf h}_m$ and $\bar {\mathbf{m}}$ for computing the solution for ${\cal P}_2$. As a result, the convex optimization problem
${\cal P}_2$ is formulated as
\begin{equation}\label{opt21}
\min_{\mathbf{h},\mathbf{\tilde m}_z}{1 \over 2} \left\| \bar {\mathbf D}
\left[\begin{array}{c} {\mathbf h} \\
{\mathbf {\bar m}}_z \end{array}\right] - {\mathbf m}^o\right\|_2^2 +
{\lambda_2} \|{\mathbf{h}}_m\|_\ast,
\end{equation}
$$\textrm{subject to:}\ \mathbf{h}_m \succeq {\mathbf 0} \ \textrm{and} \
{\mathbf 0}\preceq \mathbf{\bar m} \preceq {\mathbf c} \odot {\bar {\mathbf{h}}_m}.$$
%With a properly selected ${\mathbf c}$, the convex hull $\mathbf{h} \succeq {\mathbf 0} \ \textrm{and}$ and ${\mathbf 0}\preceq \mathbf{\tilde m}_z \preceq {\mathbf c} \odot {\bar {\mathbf{h}}}$ form a
%contains the true solutions for $\mathbf{h}_m$ and $\mathbf{\bar m}$.
The above convex optimization problem (\ref{opt21}) can be equivalently formulated as
\begin{equation}\label{opt22}
\min_{\mathbf{h},\mathbf{\bar m},\mathbf{q}}
{1 \over 2} \left\| \bar {\mathbf D}
\left[\begin{array}{c} {\mathbf h} \\
{\mathbf {\bar m}} \end{array}\right] - \bar {\mathbf m}^o\right\|_2^2 +
{\lambda_2} \|{\mathbf{h}}_m\|_\ast +
{\gamma_2} \|{\mathbf c} \odot \mathbf{\bar h}-\mathbf{\bar m}-\mathbf{q}\|_2^2,
\end{equation}
$$\textrm{subject to:}\ \mathbf{h}_m \succeq {\mathbf 0}, \ \mathbf{\bar m} \succeq {\mathbf 0} \ \textrm{and} \
{\mathbf {q}}\succeq \mathbf{0},$$
where $\lambda_2$ and $\gamma_2$ are regularization parameters.

\section{Phantom and in-vivo data experiments}
\subsection{Cartesian and Non-Cartesian Data}
%The proposed two-step convex optimization algorithm is applied to a set of scanned phantom data used for experiments in \cite{lustig_2010}, which is available at
%http://www.eecs.berkeley.edu/mlustig.
%The phantom data set was scanned on a GE Signa-Excite 1.5T scanner using a 4-channel cardiac coil set with a spiral gradient echo sequence.
%The spiral trajectory was designed with 60 interleaves, 30 cm field of view and 0.75 mm in-plane resolution and the readout time was 5 ms. The $k$-space Data was undersampled by choosing 20 out of 60 interleaves. .
%For image reconstruction in case of non-cartesian data sets, the NUFFT code by \cite{fessler_2003} was applied.

Two sets of in-vivo MRI data
were adopted to test the proposed convex reconstruction algorithm. The first is a brain data set (available at \url{http://black.bme.ntu.edu.tw/tool\_sense.html}) of a healthy human volunteer was acquired by a 3 Tesla SIEMENS Trio scanner with an eight-channel head array and an MPRAGE (3D Flash with IR prep.) sequence. The parameters of the scan were $TR/TE =2530/3.45$ ms, $TI = 1100 ms$, $N^2 =256\times 256$, flip angle $= 7^{\circ}$, slice Thickness $=1.33$ mm and $FOV =256\times 256$ mm$^2$. The fully acquired $k$-space data in the cartesian co-ordinate system are manually undersampled to obtain the uniform sampling with additional auto-calibration signal (USACS) patterns of $4$-, $8$-, $12$- and $16$-fold acceleration rates, respectively, with additional 36 extra auto-calibrating signal (ACS) lines in the central $k$ space region along the phase encoding direction in each pattern. As a result, the corresponding net reduction factors are $f_{net}=2.56, \ 3.76, \ 4.49$, and $4.92$, respectively.

%Two pseudo-random sampling schemes were further applied for undersampling the full in-vivo data set. One is a pseudo-angular pattern designed to select $18.80\%$ of all $k$-space data, as shown in Fig.\ref{a4},
%and the other is a pseudo-spiral pattern which contains $28.81\%$ of the whole data points, as given in Fig.\ref{s4}.

%The second data set (available at \url{http://black.bme.ntu.edu.tw/tool\_sense.html}) was acquired on a 3 Tesla SIEMENS Trio scanner from a healthy human volunteer using an eight-channel head array and MPRAGE (3D Flash with IR prep.) sequence with parameters $TR/TE =2530/3.45$ ms, $TI = 1100 ms$, $N^2 =256\times 256$, Flip angle $= 7^{\circ}$, slice Thickness $=1.33$ mm and $FOV =256\times 256$ mm$^2$.
%The data acquisition was performed in the cartesian coordinate system
%and the fully acquired $k$-space data were manually undersampled by USACS sampling schemes along the phase
%encoding direction by a nominal data reduction factor $f_{nom}=4$ with 36 ACS lines around the center region, corresponding to a net reduction factor $f_{net}=2.56$. The undersampled data sets for reduction factors $f_{nom}=8,\ 12, \ 16$, corresponding to $f_{net}=3.76, \ 4.49, \ 4.92$, respectively, of the same USACS pattern were also obtained.

Another $in-vivo$ data set of spine (available at \url{http://ece.tamu.edu/jimji/pulsarweb/downloads.htm}) was acquired from a $4$-channel cervical-thoracic-lumbar spine array using a fast spoiled gradient-echo sequence and parameters $TR/TE =300/12$ ms, $RBW = 62.5 kHz$,  $N^2 =256\times 256$, tip angle $= 15^{\circ}$ and $FOV= 32\times 32$ cm$^2$. The fully acquired $k$-space data in the cartesian coordinate system are undersampled to generate the USACS patterns of $4$-, $6$- and $8$-fold acceleration rates, respectively, with each pattern having $32$ extra ACS lines along the phase encoding direction.

The proposed two-step convex optimization algorithm is also applied to a set of scanned non-cartesian phantom data, which is available at
\url{http://www.eecs.berkeley.edu/\~mlustig}.
The phantom data set was scanned on a GE Signa-Excite 1.5T scanner using a $5$-channel cardiac coil set with a spiral gradient echo sequence.
The spiral trajectory was designed with $60$ interleaves, $30$ cm field of view, $0.75$ mm in-plane resolution and  readout time of $5$ ms. The $k$-space Data was undersampled by choosing $20$ out of $60$ interleaves. For image reconstruction in case of non-cartesian data sets, the NUFFT code by \cite{fessler_2003} was applied.

\begin{figure}[!t]
\centerline
{\includegraphics[width=3in]{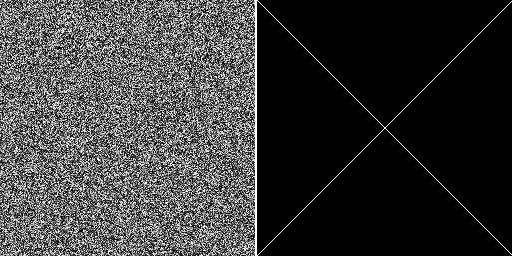}}
\caption{Two examples of initial images}
\label{f1}
\end{figure}

\begin{figure}[!t]
\centerline
{\includegraphics[width=3in]{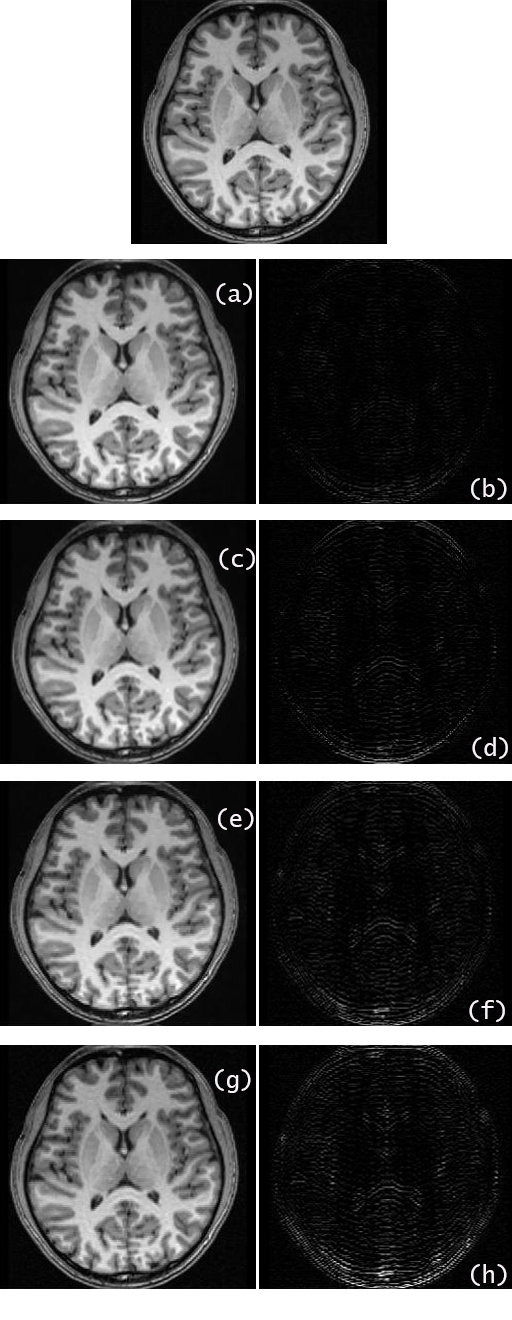}}
\caption{Images reconstructed by our method for $8$ channel Brain Data,corresponding error images have been shown adjointly, Fig (a), (c), (e) and (g) resulted by a nominal reduction factor of $4, \ 8, \ 12$ and $16$ respectively and Fig (b), (d), (f) and (h)  are the corresponding error images.}
\label{fig3}
\end{figure}

\subsection{Computational set ups}

The computation of the split-bregman based optimization problem (\ref{opt11}) was implemented using the iterative algorithm proposed in \cite{goldstein_2009}. The algorithm proposed in \cite{majumdar_2011}, together with the result in \cite{law_1974}, was adopted for resolving the nuclear norm regularized optimization problem (\ref{opt21}). LSQR \cite{law_1974} tools were used in nuclear norm regularized optmization. For wavelet transformations in the $l_1$
Wavelet regularized reconstruction, David Donoho's Wavlab codes\cite{donoho_1996} were used. Two wavelet familes, "Haar" and "Daubechies" were selected as the sparsifying transform basis. The regularization parameters were empirically chosen and a tolerance value of $10^{-6}$ is selected for each step of iteration. Both algorithms are programmed with Matlab (Math-Works, Natick, MA, USA).

To evaluate the reconstruction accuracy, the reconstructed images, denoted by $\mathbf h^o$, are compared with the sum of square (SOS) image, denoted as $\mathbf h_{SOS}$, which is reconstructed using the fully sampled $k$-space data. The the normalized mean square error (NMSE) of $\mathbf h^o$ is defined as
$$e_{NMSE}={{\|\mathbf h^o-\mathbf h_{SOS}\|^2}\over {\|\mathbf h_{SOS}\|^2}}.$$

The reconstructed images by the proposed algorithm are computed and compared with the reconstructed images by conjugate gradient (CG)-SPIRiT \cite{lustig_2010} with $l_1$ penalty, GRAPPA \cite{griswold_2002}, JSENSE\cite{leslie_2007} and IRGN-TGV \cite{knoll_2012} algorithms for the in-vivo $8$ channel brain data sets under the same data reduction conditions. The Matlab codes as well as the regularization parameters and initial conditions, where applicable, for computations of these algorithms are originated from \url{http://www.eecs.berkeley.edu/~mlustig/Software.html} for GRAPPA and CG-SPIRiT,
\url{http://cai2r.net/sites/default/files/software/irgntv.zip} for IRGN-TGV and \url{https://pantherfile.uwm.edu/leiying/www/index_files/software.htm} for JSENSE.

%\begin{figure}[!htb]
%\centerline
%{\hspace{-0.05\linewidth}\includegraphics[width=4.8in]{brain2.jpg}}
%\caption{Simulation Results for $32$ channel compressed Brain Data with our reconstruction method, Fig (a), (b) and (c) correspond to a nominal reduction factor of $4, \ 8$ and $12$ respectively, Fig (d)-(f) depict the respective error images.}
%\label{fig2}
%\end{figure}
The global solutions of the proposed convex optimization algorithm are tested with different initial image conditions. Two typical initial image conditions, a randomly generated matrix and diagonal lines matrix, shown in Fig. \ref{f1}, of compatible dimensions with reconstructed images.

\begin{figure*}[!htb]
\centerline
{\includegraphics[width=6in]{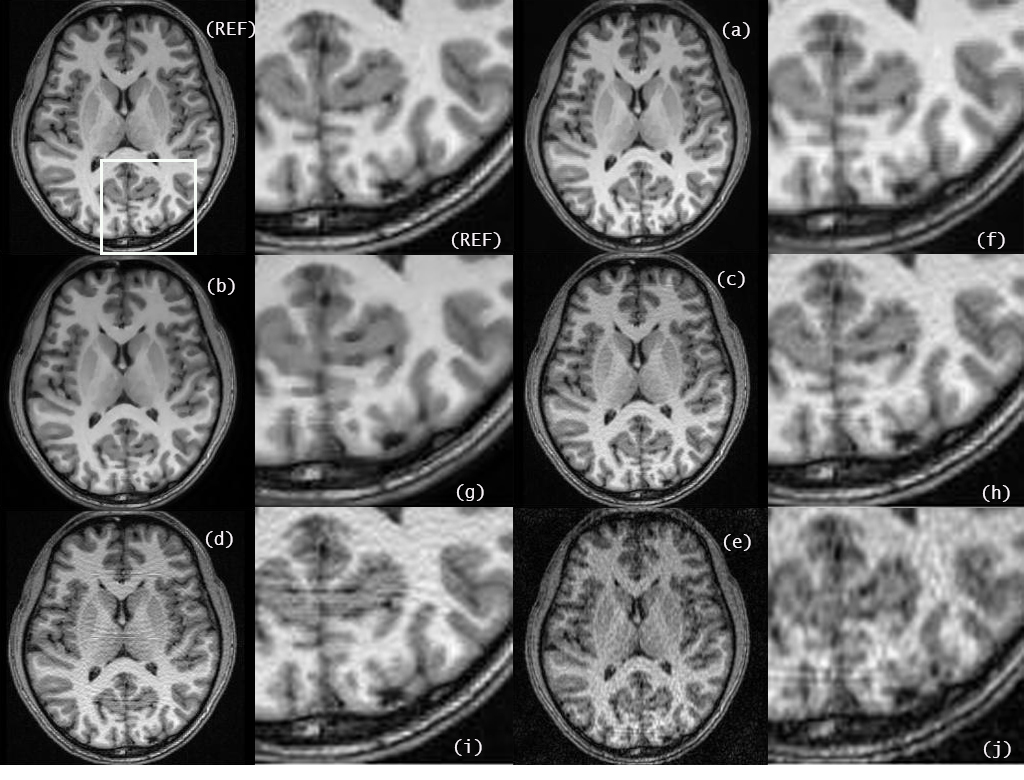}}
\caption{Comparison between reconstructed images by different algorithms, nominal reduction factor is $f_{nom}=8$, Figures (a)-(e) represent the reconstructed images by the proposed method, IRGN-TGV,  CG-SPIRiT, GRAPPA and JSENSE, respectively. A selected area of the image is zoomed for comparison and shown in the corresponding Figures (f)-(j).}
\label{fig5}
\end{figure*}
%\newpage

\subsection{In-vivo cartesian reconstruction}
Fig.\ref{fig3} (a), (c), (e) and (g) display the reconstructed images from manually under-sampled data of acceleration factors $4, \ 8, \ 12$ and $16$, respectively, for the eight-channel brain image in comparison with the reference image reconstructed by SOS from the full $k$-space data set. The regularization parameters are selected as $\lambda_1=0.01$, $\gamma_1=1$, $\lambda_2=50$ and $\gamma_2=0.05$, using the "Haar" transform.  At a smaller acceleration rate such as $f_{nom}=4$, the reconstructed image portrays good quality with very small difference from the reference image. Some quality degradation can be observed from the reconstructed images with higher acceleration factor such as $f_{nom}=16$. The NMSEs of the images are computed as $0.0027$, $0.0040$, $0.0052$ and $0.0067$ for the acceleration factors of $4, \ 8, \ 12$ and $16$, and the corresponding error images are shown in Fig.\ref{fig3} (b), (d), (f) and (h), respectively.

{\small
\begin{table}[!t]
\renewcommand{\arraystretch}{1}
\caption{Comparison between NMSE of different algorithms}
\label{trf}
\centering
\begin{tabular}{|l|r|r|r|r|r|}
\hline
\bfseries $f_{nom}$  & \bfseries SPIRiT & \bfseries GRAPPA & \bfseries  JSENSE & \bfseries  IRGN & \bfseries Our Method\\
\hline\hline
4   & 0.0032   & 0.0064   & 0.0072 & 0.0036    & 0.0027\\
8   & 0.0049   & 0.0102   & 0.0096 & 0.0048    & 0.0040\\
12  & 0.0068   & 0.0125   & 0.0120 & 0.0065    & 0.0052\\
\hline
\end{tabular}
\end{table}}

\begin{figure}[!t]
\centerline
{\includegraphics[width=2.9in]{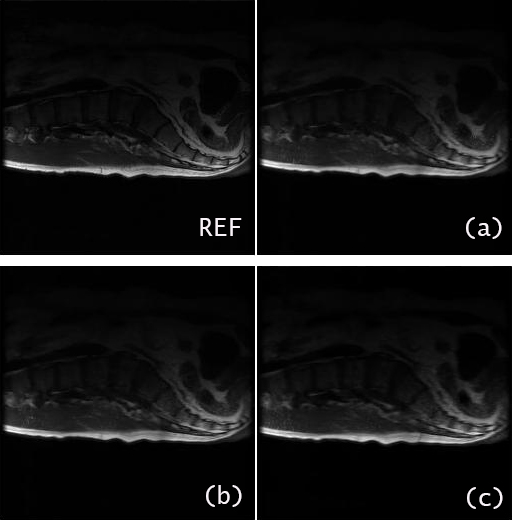}}
\caption{Reconstruction images of the spine data set by the proposed method. Fig.\ref{fig6} (a)-(c) are resulted from reduction rates $f_{nom}=$$4, 6$ and $8$ respectively.}
\label{fig6}
\end{figure}

\begin{figure}[!t]
\centerline
{\includegraphics[width=3.5in]{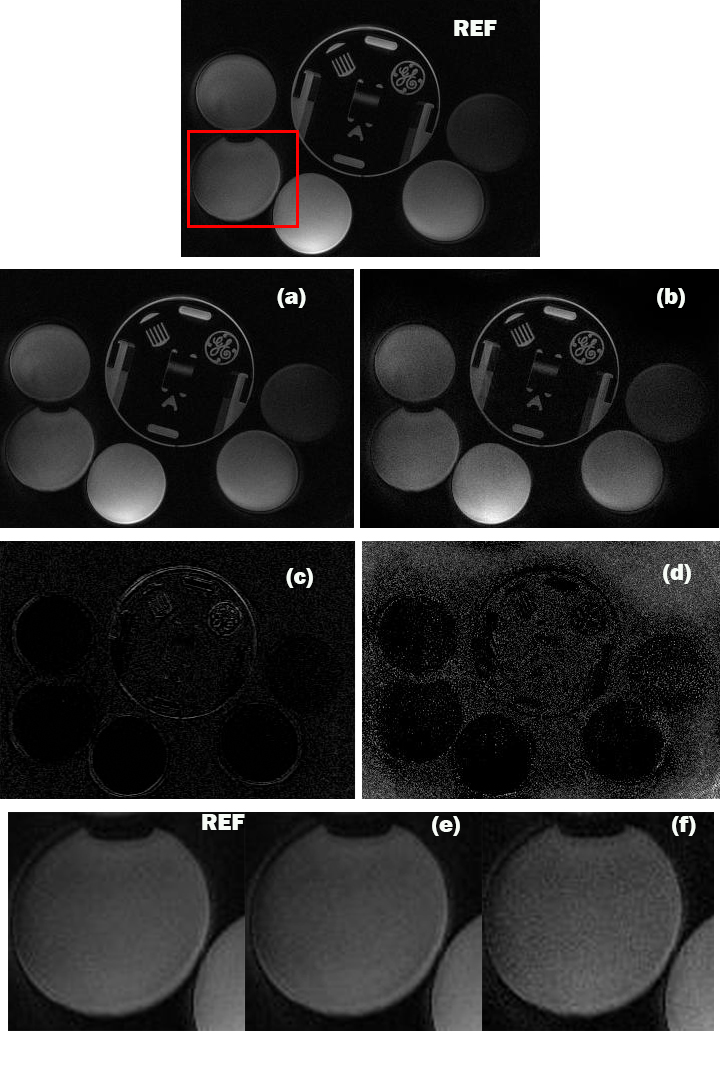}}
\caption{Reconstruction results of the spiral non-cartesian data. Fig.\ref{fig7} (a) and (b) are the reconstructed image by our method and CG-SPIRiT and Fig.\ref{fig7} (c) and (d) are their error images in comparison with the reference image, respectively. Fig.\ref{fig7} (e) and (f) display a zoomed portion of the images reconstructed by our method and CG-SPIRiT, respectively, in comparison with the same portion of the reference image.}
\label{fig7}
\end{figure}

For the brain date set of acceleration factor $8$, Fig.\ref{fig5} presents the reconstruction image of the proposed method in comparison with the images reconstructed by other commonly known algorithms, which are IRGN-TGV, CG-SPIRiT, GRAPPA and JSENSE. A selected area, as blocked within the marked rectangle in the reference image of Fig.\ref{fig5}, is zoomed for each reconstructed image and jointly displayed with the corresponding full size image for ease of comparison. Among these algorithms, GRAPPA and CG-SPIRiT are members of the second group using the SOS operation and IRGN-TGV and JSENSE are nonlinear optimization algorithms of the third group. Noticeable errors and artifacts are shown in reconstruction results of Fig.\ref{fig5} (c), (d) and (e) of CG-SPIRiT, GRAPPA and JSENSE algorithms, respectively. More careful observation can also find artifacts in the zoomed image Fig.\ref{fig5} (g) reconstructed by the nonlinear iterative algorithm IRGN-TGV. The NMSE values of the reconstructed images by the different algorithms are listed in Table \ref{trf}. The the average computational time durations of repeatedly running the original Matlab codes of the different reconstruction algorithms on a workstation with Intel Xeon Processor E5-2609 and 16 GB RAM are given in Table \ref{trc}.

%from each of the reconstructed images has been selected and zoomed for ease of comparison, they are illustrated jointly with their full resolution images. CG-SPIRiT and GRAPPA utilize the sum-of-square (SOS) reconstruction for the image reconstruction in the last step whereas our algorithm uses nuclear norm-regularized linear optimization. Superiority of our algorithm is observed with comparison to CG-SPIRiT and GRAPPA as both display noticeable artifacts. The joint sensitivity and image reconstruction method JSENSE sustains much noise although other extensions related to this algorithm i.e. Sparse-SENSE \cite{liu_2008}, CS-SENSE \cite{liang_2009} and self feeding Sparse-SENSE \cite{huang_2010} may improve the result a bit, also it should be noted that most of these sensitivity-based methods heavily rely on pre-estimation of sensitivity maps and joint estimation leads to local only solutions.

%Although IRGN-based TGV penalized method shows noise free smooth image but with few noticable artifacts specially in the selected area compare to the result by our method, it has been tested that reducing TV regularization parameter resulted a increased detail as well as artifacts for IRGN with comparison to our method, also it should be remembered that the algorithm is non-linear. Comparison between estimated $e_{NMSE}$'s are presented in table \ref{trf} for a nominal reduction rate of $4, \ 8$ and $12$ with $36$ extra ACS lines around the center along the phase encoding direction..

For the in-vivo spine data set, the regularization parameters of the proposed algorithm are chosen as $\lambda_1=0.05$, $\gamma_1=0.5$, $\lambda_2=10$ and $\gamma_2=0.1$. The wavelet transform matrix is ${\mathbf{W}}=$"Haar". The reference image constructed by the full data set and SOS operation is given in Fig.\ref{fig6}, followed that the reconstructed images by the proposed method, for nominal undersampling rate $f_{nom}=4$ in Fig.\ref{fig6}(a). The estimated NMSE
of this reconstructed image is $e_{NMSE}=0.0038$. For higher reduction rates of $f_{nom}=6$ and $f_{nom}=8$, the corresponding reconstructed images by the proposed algorithm are shown in Fig.\ref{fig6}(b) and \ref{fig6}(c), with estimated NMSE values of $0.0044$ and $0.0049$, respectively.

\begin{table}[!htb]
\renewcommand{\arraystretch}{1}
\caption{Average computational times (secs) of different algorithms
over six repeated reconstruction processes}
\label{trc}
\centering
\begin{tabular}{|c|c|c|c|c|}
\hline
%\bfseries iterations  &
\bfseries  SPIRiT & \bfseries GRAPPA & \bfseries  JSENSE & \bfseries  IRGN & \bfseries Our Method\\
\hline\hline
%6   &
23   & 31   & 221 & 228    & 64\\
\hline
\end{tabular}
\end{table}

\subsection{Non-cartesian data reconstruction of phantom}
For the phantom data of the spiral pattern, a reference image as given in Fig.\ref{fig7} is produced by applying the NUFFT and SOS operations on the full data set.
The proposed algorithm, with regularization parameters $\lambda_1=0.01$, $\gamma_1=0.001$, $\lambda_2=10$, $\gamma_2=0.5$ and the wavelet transform matrix ${\mathbf{W}}=$"Daubechies", is applied to the undersampled spiral phantom data and produces the reconstructed image in Fig.\ref{fig7}(a) and the corresponding error image, with respect to the reference image, in Fig.\ref{fig7}(c). Another algorithm CG-SPIRiT which is capable of non-cartesian reconstruction is also applied to the same undersampled data set, resulting in the reconstructed image
in Fig.\ref{fig7}(b) and the corresponding error image in Fig.\ref{fig7}(d). A selected area from both reconstruction results are cropped and scaled up, as shown in Fig.\ref{fig7} (e) and (f) respectively, to demonstrate the effectiveness of our proposed algorithm.
%The results demonstrate effectiveness of the proposed reconstruction algorithm.
%\newpage
\section{Discussions}
The proposed convex optimization approach to pMRI reconstruction is build upon a 
convex solution space which exists only for the magnitude image function but does not exist for any real and complex valued images. This paper formulated a two-step convex optimization to solve the pMRI reconstruction and it is possible to solve the convex optimization problem with alternative formulations. 

The solution space of the proposed two-step optimization is a convex hull specified by the constraints in (\ref{opt21}) which contains the true solution for the image and sensitivity functions. In general, the optimal solution and its computation depend on selections of the regularization parameters as well as the constraint vector $\mathbf c$ in (\ref{opt21}). A priori knowledge of the image and sensitivity functions and empirical tests of the parameter and constraints can be helpful for efficient and accurate computing of the solution. The proposed algorithm produces a global solution in the sense that the solution is unique and independent of the initial
image value for the computational algorithm. This is a distinctive characteristic of the proposed method, because all other existing methods for optimization of the image reconstruction, without using previously estimated sensitivity values or the SOS operation, can provide only local solutions, which are dependent on the initial value of the algorithm. In the phantom and in-vivo data reconstructions by the proposed method, all global solutions of the proposed algorithms were tested with different initial conditions including that shown in Fig.\ref{f1} and their uniqueness was verified. In contrast, the solutions of other algorithms based on non-convex optimization, such as IRGN-TGV and JSENSE, are local only. Our experiments showed that their reconstruction results are very different subject to different initial conditions.

The experiments of the in-vivo and phantom data sets demonstrated better image reconstruction quality of the proposed method than that of GRAPPA and CG-SPIRiT which are algorithms of the second group using SOS operation. It indicates that the optimization with properly specified regularization terms can provide better reconstruction results than that of the simple SOS operation. This reconstruction improvement, however, involves more workload in the iterative computing of the optimal solution, which can be seen from Table\ref{trc} of the computational time durations. Because of the linear and convex nature of our proposed algorithm, it has faster and more efficient computation of the optimal solutions in comparison with the nonlinear optimization algorithms IRGN-TGV and JSENSE as shown by their computational time durations in Table\ref{trc}.

%The total simulation duration was approximately in between 63-66 seconds for in-vivo brain dataset for gamma_1=0.005 in a Intel? Xeon? Processor E5-2609 with 16 GB RAM workstation, increasing gamma_1 would decrease the running time although some visible noise will tend to exhibit for larger gamma_1. Attempting to negotiate tolerance will also cause lower running time as fewer iterations needed for the algorithm to converge to the optimal solution.

The proposed algorithm for computing the phantom and in-vivo images is only a specific realization of the general linear and convex optimization method in terms of the two-step
optimization problems ${\cal P}_1$ and ${\cal P}_2$. Algorithms using other regularization terms and realizations
relevant to different reconstruction requirements can be possible and will be subject to future studies.

\section{Conclusion}
The reconstruction of MR images based on undersampled $k$-space data by optimization methods for pMRI is known as a nonlinear and nonconvex problem. It is a recently active research area in MRI
reconstruction and the existing optimization methods without using estimated sensitivity functions or the SOS operation can only provide local but not global solutions. And the solutions for such a nonlinear and nonconvex problem involve complicated computational procedures and iterations. In this paper, a linear equation is derived for the undersampled $k$-space data set in terms of the magnitude image function, which enables the formulation of the pMRI reconstruction into a two-step convex optimization problem. It is applicable to both cartesian and non-cartesian data sets and can provide a globally optimal solution and faster computation for the pMRI reconstruction problem. An algorithm is presented in this paper and applied to phantom and in-vivo MRI data sets to demonstrate the reconstruction performance and effectiveness of the proposed method.

%% use section* for acknowledgement
%\section*{Acknowledgment}
%% Can use something like this to put references on a page
%% by themselves when using endfloat and the captionsoff option.
%\ifCLASSOPTIONcaptionsoff
%  \newpage
%\fi



%\bibliographystyle{IEEEtran}
%\begin{thebibliography}{1}
%
%\bibitem{roemer_1990}
%\end{thebibliography}
%\newpage
\bibliographystyle{IEEEtran}
\bibliography{ref2}



%\begin{IEEEbiography}{Professor Cishen Zhang}
%\end{IEEEbiography}

%\begin{IEEEbiographynophoto}{Ifat-Al Baqee}

%\newpage


%\vfill


\end{document}